\documentstyle[pra,aps,preprint]{revtex}

\def\beq{\begin{equation}}
\def\eeq{\end{equation}}

\def\reff#1{(\ref{#1})}

\def\vekt#1{\bbox{#1}}

\def\vektr{\vekt{r}}

\def\vektE{\vekt{E}}

\def\vektnabla{\vekt{\nabla}}

\def\halb{\frac{1}{2}}

\def\Edach{\hat{E}}
\def\Ecrit{E_{\mbox{\rm\scriptsize crit}}}

\def\landaurate{W_{{\mbox{\rm\scriptsize L}}}}
\def\keldyshrate{W_{{\mbox{\rm\scriptsize K}}}}
\def\adkrate{W_{{\mbox{\rm\scriptsize ADK}}}}
\def\posthrate{W_{{\mbox{\rm\scriptsize cl}}}}
\def\krainovrate{W_{{\mbox{\rm\scriptsize Kr}}}}
\def\mulserrate{W_{{\mbox{\rm\scriptsize Mu}}}}
\def\Icl{I_{{\mbox{\rm\scriptsize cl}}}}

\def\energy{{\cal{E}}}

\def\pabl#1#2{\frac{\partial #1}{\partial #2}}

\def\imagi{\mbox{\rm i}}

\begin{document}
\draft

\title{Exact field ionization rates in the barrier suppression-regime\\ from
numerical TDSE calculations}
\author{D.~Bauer and P.~Mulser} 
\address{Theoretical Quantum Electronics (TQE)\cite{www}, Darmstadt
University of Technology,\\ Hochschulstr.\ 4A, D-64289 Darmstadt, Germany}

\date{\today}

\maketitle

\begin{abstract}
Numerically determined ionization rates for the field ionization of
atomic hydrogen in strong and short laser pulses 
are presented. The laser pulse intensity reaches the so-called
``barrier suppression ionization'' regime where field ionization
occurs within a few half laser cycles. 
Comparison of our numerical
results with analytical theories frequently used shows poor agreement. 
An empirical formula for the ``barrier suppression ionization''-rate
is presented. This rate reproduces very well the course of the numerically
determined ground state populations for laser pulses with different
length, shape, amplitude, and frequency.    
\end{abstract}


\section{Introduction}
With the ``table-top'' laser systems, nowadays available,  laser pulse
peak field strengths
much greater than the binding field of the outer atomic electrons can
be achieved (see e.g.\ \cite{natoI} for an overview). Above a certain
threshold electric field the electron is able to escape even
classically from the atomic nucleus, i.e., without tunneling through
the barrier formed by the Coulomb potential and the external electric
(laser) field. This regime is called 
``barrier suppression ionization'' (BSI) \cite{augst}. 
 
In combination with the dramatic progress
in decreasing the pulse duration below 10~fs  \cite{zhoui,stingl,zhouii,barty} 
new features in the ionization
dynamics are expected. In particular, ionization at such high field
strengths occurs mainly within a
few half laser cycles, i.e., on a sub-femtosecond time scale, provided
that the pulse rises fast enough so that tunneling
contributes negligibly to the overall ionization.
Fast depletion of  bound states within
one half laser cycle leads to a non-isotropic electron
distribution. Apart from the peaked angular distribution of the photo
electrons in electric field
direction, in the BSI-case there is
also an asymmetry along this field axis \cite{bauerthesis}. 
This opens up the possibility to manipulate the
electron distribution function of laser produced plasmas.
By ``tailoring'' the
pulse shape the plasma formation process may be controlled according
to the application under consideration, e.g., harmonics generation
\cite{kulander_harm}, or XUV laser schemes \cite{fill}. 

Experimentally observed ion yields are usually analyzed by means
of tunneling theories among these Ammosov-Delone-Krainov (ADK) \cite{adk}, Keldysh
\cite{keldysh}, Keldysh-Faisal-Reiss (KFR) \cite{reiss}
or Landau \cite{landau} theory are the most prominent ones. However, it is, in
general, not possible to get good agreement for several ion species
without ``shifting'' the laser intensity \cite{augst}. 
By examining the derivations of KFR-type theories it becomes obvious
that they {\em should} fail in the ``barrier suppression ionization''
(BSI) regime because the transition between an {\em unperturbed} initial state and a Volkov state is calculated there. However, the influence of the strong laser field on the {\em inneratomic} dynamics must not be neglected in BSI. 
An attempt to extend the ADK-theory to BSI has been undertaken
\cite{krainov}. A pure classical ionization rate has been proposed 
recently \cite{posth}.

In this paper we compare numerically determined ionization rates for
various kinds of pulse shapes and peak field strengths with
results predicted by several analytical derivations: 
the Landau tunneling formula \cite{landau}, the Keldysh rate
\cite{keldysh}, the ADK formula \cite{adk} and its extension to the
BSI-regime \cite{krainov}, a classical rate derived by Posthumus {\em
et al.}\ \cite{posth} and a tunneling rate suggested by Mulser
\cite{mulser}. In our numerical studies we restrict ourselves to the
ionization of atomic hydrogen in
an intense, short, linearly polarized laser pulse. We focus on the
field strength region where the ionization rate is of the order of the
laser frequency because ionization occurs within a few half
laser cycles in this case.

In Section \ref{theory} we review the time-dependent Schr\"odinger
equation (TDSE) of field ionization. Moreover, we state the analytical
formulas used for comparison
with our numerical results. In Section \ref{numericalresults} we
present our numerical results for various pulse shapes and field
strengths. The numerical results are discussed in Section
\ref{disc}.
We conclude in Section \ref{conclusion}. Details on the
numerical method are attached in the Appendix.

\section{Theory} \label{theory}
\subsection{Time-dependent Schr\"odinger equation (TDSE)}
The TDSE for an
electron interacting with the nuclear potential $-Z/r$ and the laser
field $\vektE(t)$ in dipole approximation and length gauge reads \cite{faisalbook}
\beq \imagi\pabl{}{t}\Psi(\vektr,t) = \left( -\frac{\vektnabla^2}{2} -
\frac{Z}{r} + \vektr\vektE(t) \right)
\Psi(\vektr,t)
 \label{schr_laengen} \eeq 
(atomic units (a.u.) are used throughout this paper \cite{au}). If the electric
field is chosen to be directed along the $z$-axis, cylindrical
coordinates are introduced and the Ansatz $\Psi(\rho,\varphi,z,t) =
\psi(\rho,z,t) \exp(\imagi m \varphi)  (2\pi)^{-1/2} $
is made,
the TDSE assumes the following two-dimensional form,
\beq \imagi \pabl{}{t} \psi = -\halb \left( \frac{1}{\rho} \pabl{}{\rho}
\Bigl(\rho \pabl{}{\rho}\Bigr) -\frac{m^2}{\rho^2} + \pabl{^2}{z^2} \right) \psi
+ (zE(t)-\frac{Z}{\sqrt{\rho^2+z^2}})\psi, \label{sg_in_psi} \eeq
and the normalization condition 
\beq \int_0^\infty d\rho\, \rho \int_{-\infty}^\infty  dz\,  \vert \psi
(\rho,z,t)\vert^2 = 1 \eeq
holds. The TDSE \reff{sg_in_psi} was
numerically solved
first by Kulander in 1987, but for intensities below
$10^{15}$~W/cm$^2$ \cite{kulander_I}.
 
In a recent work by Kono {\em et al.}\ \cite{kono} it was systematically examined
for what parameter $\lambda$ the substitution 
\beq \Phi(\xi,z,t)=\sqrt{\lambda} \xi^{\lambda-1/2}
\psi(\xi^\lambda,z,t), \qquad \xi^\lambda=\rho,\quad z=z, \quad t=t,
 \label{substi} \eeq
is most favorable numerically. It turned out that the choice $\lambda=3/2$
 is best, both for stability and accuracy. The TDSE
corresponding to the substitution \reff{substi} is given in
Appendix \ref{app}. We used a Peaceman-Rachford scheme to propagate the wavefunction $\Phi(\xi,z,t)$ 
(see Appendix \ref{app} or Ref.\ \cite{kono} for
details). Absorbing boundary conditions were implemented which keep
the main interaction region in the vicinity of the atomic nucleus free
from otherwise reflected probability density.

In all our calculations we started from the 1s ground state, i.e.,
$m=0$. The stable ground state on the numerical grid (which is
slightly different from the analytical solution of the
Coulomb problem, depending on the grid-spacing ) 
was determined by applying our propagation scheme
with an imaginary timestep to the grid representation of the known
analytic solution.

\subsection{Ionization rate formulas} \label{ionirates}
In this Section we review the ionization rate formulas used for
comparison with our numerical results of Section
\ref{numericalresults}. 
If we assume that an ionization rate $W[E(t)]$ is given, the
probability for the electron to remain bound is
\beq \Gamma(t)=\exp\left(-\int_0^t W [ E(t')]\, dt' \right) . \eeq
We take 
\beq \Lambda(t)=1-\Gamma(t) \label{ionirate} \eeq 
as the ionization probability which is, apart from a small time-shift,
equivalent to the common procedure to calculate the amount of
probability to find the electron in a small volume around the atomic nucleus,
\beq \Lambda'(t)=1-\int_0^a d\rho\, \rho \int_{-a}^a  dz\,  \vert \psi
(\rho,z,t)\vert^2, \qquad a\approx 5\,\mbox{\rm a.u.} \eeq 
We assume that the laser pulse ``hits'' the atom at $t=0$ (or ionization is
negligible for $t<0$). 

\subsubsection{Landau formula}
Landau \& Lifshitz derived a formula for the ionization rate of
hydrogen when the electron is in the ground state initially \cite{landau}. The
result is easily extended to hydrogen-like ions (where the ground
state energy is $\energy_0=-Z^2/2$), 
\beq \landaurate = 4\frac{(2\vert\energy_0\vert )^{5/2}}{E} \exp\left(
-\frac{2(2\vert\energy_0\vert )^{3/2}}{3E} \right)
. \label{landaurate} \eeq

\subsubsection{Keldysh formula}
Keldysh perturbatively calculated the transition rate from an initial
bound state to a state representing a free electron in a laser field
(Volkov state) \cite{keldysh},
\beq 
\keldyshrate= \frac{(6\pi)^{1/2}}{2^{5/4}}\energy_0 \left(
\frac{E}{(2\energy_0)^{3/2}} \right)^{1/2} \exp\left(
-\frac{2(2\vert\energy_0\vert )^{3/2}}{3E} \right). \eeq

\subsubsection{Ammosov-Delone-Krainov (ADK) formula}
Ammosov, Delone, and Krainov derived a tunneling ionization
rate for complex atoms in an ac electric field \cite{adk}. The initial 
state is described by an effective quantum number $n^*$ and the
angular and magnetic quantum numbers $\ell$ and $m$, respectively. The ADK-result
reads
\beq \adkrate = C_{n^*\ell}^2 f(\ell,m) \vert\energy_0\vert \left(
\frac{ 3E}{\pi (2\vert \energy_0\vert)^{3/2}} \right)^{1/2} \left(
\frac{2}{E} (2\vert \energy_0\vert )^{3/2}\right)^{2n^*-\vert m\vert
-1} \exp \left(
-\frac{2(2\vert\energy_0\vert )^{3/2}}{3E} \right) \label{adkrate}
\eeq
with
\[ C_{n^*\ell}=\left( \frac{2\mbox{\rm e}}{n^*} \right)^{n^*} (2\pi
n^*)^{-1/2}, \qquad f(\ell,m) = \frac{ (2\ell+1)(\ell +\vert
m\vert)!}{2^{\vert m\vert } \vert m\vert ! (\ell -\vert m \vert
)!}. \]
The constant e in the coefficient $C_{n^*\ell}$ is Euler's number
2.71828... In the derivation of the ADK rate \reff{adkrate} 
averaging over one laser cycle was performed. The validity of the
ADK-formula is expected to be best for $n^* \gg 1$, $E\ll 1$, and
$\omega \ll \vert\energy_0\vert $. 

\subsubsection{BSI extension to ADK}
Krainov suggested an extension of ADK-theory to incorporate BSI
\cite{krainov}. The result is
\beq \krainovrate = \frac{4\sqrt{3}}{\pi n^*} \frac{E}{(2E)^{1/3}}
\left( \frac{4\mbox{\rm e} (\vert\energy_0\vert
)^{3/2}}{En^*}\right)^{2n^*} \int_0^{\infty} \mbox{\rm
Ai}^2\left(x^2+\frac{2\vert\energy_0\vert}{(2E)^{3/2}} \right) x^2\,
dx \label{krainovrate} \eeq
where Ai denotes the Airy-function. Formula \reff{krainovrate} reduces
to the usual ADK rate \reff{adkrate} in the limit of a relatively weak
laser field (tunneling limit).

\subsubsection{Classical rate proposed by Posthumus et al.}
Recently, Posthumus and co-workers proposed a purely classical BSI
ionization rate \cite{posth}. Taking the equipotential surface corresponding to the
atomic ground state and examining its intersection with the
field-deformed Coulomb potential enables the authors to calculate the
rate from a geometrical viewpoint. Their result reads
\beq \posthrate=\frac{1-\energy_0^2/(4 Z E)}{2 T_0},\qquad T_0=
\frac{\pi Z}{\vert \energy_0\vert
(2\vert\energy_0\vert)^{1/2}}. \label{posthrate} \eeq
$T_0$ is the classical orbit period for the so-called ``free
falling''-trajectories with zero angular momentum. 
The authors of \cite{posth} 
present also a cycle-averaged expression of the
rate. They finally suggest to take $\posthrate + \adkrate(\Icl)$ as the
total ionization rate in the BSI regime where $\Icl$ is an appropriate
threshold intensity.

\subsubsection{Tunneling rate proposed by Mulser}
Mulser calculated the ionization rate by approximating the tunneling
barrier formed by the Coulomb potential and the external field with a barrier 
parabolic in shape \cite{mulser}. 
After calculating the transmission coefficient
through this parabolic barrier and making an  assumption for the tunneling current the rate formula
\beq \mulserrate = \frac{\vert \energy_0\vert}{\vert\beta\vert} \ln
\frac{A+\exp\vert\beta\vert}{A+1},\qquad \mbox{\rm where}  \qquad
A=\exp\left( -\frac{7-3\alpha}{4}C\right), \label{mulserrate} \eeq
\[ \beta=\frac{3+\alpha}{4} C, \qquad \alpha=\frac{4
E^{1/2}}{(2\vert\energy_0\vert )^{3/4}}, \qquad C=-\pi (2\vert\energy_0\vert)^{1/8}
\frac{2\vert\energy_0\vert}{2^{1/2}E^{3/4}},  \]
is obtained. 

\section{Numerical results} \label{numericalresults}
In this Section we study the ionization dynamics of the 1s atomic
hydrogen electron under the influence of the external laser field
$E(t)$. The laser field is assumed to have the form
\beq E(t)=\Edach(t)\sin(\omega t +\varphi) \eeq
where $\Edach(t)$ is the pulse shape-function and $\varphi$ is a
constant phase. In the following we vary the pulse envelope
$\Edach(t)$, the laser frequency $\omega$, and the phase $\varphi$ in order to examine their
influence on the temporal evolution of the ground state probability
$\Gamma(t)$.
 
\subsection{Instantaneously switched on dc field}
Although the dc field instantaneously switched on is, from the
experimental point of view, not realistic at all, this case delivers
useful insight in how important transient effects might
be. Furthermore it is interesting to check whether ionization occurs 
with a constant rate after transient effects have died out. 

In the instantaneously switched on field-case the envelope function is
\beq \Edach(t)=\Edach=\mbox{const.} \qquad \mbox{\rm for} \qquad t>0 \qquad (0 \quad
\mbox{\rm otherwise}). \eeq

In Figure \ref{fig_one} the ground state population $\Gamma(t)$ is
plotted vs time for the five different amplitudes
$\Edach=0.1,0.2,0.3,0.4$ and $0.5$. One easily verifies that after a
very short transient period of about $2$~a.u.$=0.048$~fs the constant
rate-behavior sets in. 
This transient time period may be estimated by purely classical
considerations if one assumes that the atomic response time is similar
to that of a classical system with an electron density corresponding
to the quantum mechanical probability density of the ground state.
The electron density then is $n_e\approx (4\pi/3)^{-1}$~a.u.\ which
leads to a ``plasma frequency'' $\omega_p \approx 3^{1/2}$~a.u. The
classical response time therefore would be about 3.6~a.u. $=0.09$~fs.  

The constant rates $W$ are given in the plot. We
postpone a comparison with the analytical rate formulas mentioned
above till Section \ref{disc}.

The probability density $\vert \psi(\xi,z)\vert^2$ 
after $15.5$ atomic time units for the
$\Edach=0.3$-case is shown in Figure \ref{fig_two}. 
Since we chose $\Edach>0$ the electron escapes in negative
$z$-direction.
Note the pronounced asymmetry and the 1s peak which does not move as a
whole; it rather persists at the Coulomb singularity.

\subsection{Square pulses and phase dependence} \label{sqp}
Now we study an ac field with a step-like envelope function,
\beq E(t)=\Edach \sin(\omega t + \varphi)\qquad \mbox{\rm for} \qquad t>0 \qquad (0 \quad
\mbox{\rm otherwise}). \eeq
In Figure \ref{fig_three} the ground state populations for the two
field amplitudes $\Edach=0.3$ and $0.5$ are shown. In each case three
different phases ($\varphi=0,\pi/4,\pi/2$) 
were chosen in order to check how strong ionization depends on phase
effects. The frequency was $\omega=0.2$ in these runs. 

During the course of one half cycle ionization is strongly phase dependent. 
In the $E(t)=\Edach\cos\omega t$-cases ionization is particularly strong in the beginning owing to the abrupt turn-on of the field, while in the 
$E(t)=\Edach\sin\omega t$-cases ionization starts smoothly. A steady state-rate, based on cycle-averaging, of course cannot resolve such details.

For $\Edach=0.3$ ionization lasts mainly two half cycles while
for $\Edach=0.5$ already after one single half cycle ionization is $>$
98\%. The more rapid ionization is, the stronger should be the
dependence of ionization on the phase $\varphi$. However, even in the
$\Edach=0.5$-case the two fields $E(t)=\Edach\sin\omega t$ and
$E(t)=\Edach\cos\omega t$ lead to the same net ionization {\em after
one half cycle}. Only if one is interested in the ionization dynamics
on time scales below one optical half cycle ionization becomes
phase-sensitive. However, even the shortest pulses nowadays available
have to cross the field strength region where the ionization rate is $\approx \omega$. Once this regime is passed there is
not much electron density left to be ionized by the stronger part of the pulse.

For the sake of illustration the probability density after one complete
optical cycle in the $E(t)=0.3\cos\omega t$-case is shown in Figure
\ref{fig_four}. Owing to rescattering of probability density at the
ionic core wave-packets have already built up. Closer examination
yields that subsequent wave packets in position space can be mapped to
subsequent wave packets in momentum space. These momentum
space-packets differ in energy by the amount of $\hbar\omega$, and
thus are the famous ``above threshold ionization'' (ATI)-peaks
(see \cite{schwengel} for a detailed analysis).

\subsection{Gaussian pulses}
A shape which resembles in a reasonable manner an experimental laser
pulse is Gaussian. We took
\beq E(t)=\Edach(t)\sin\omega t, \qquad \Edach(t)=\Edach \exp\left( -
\frac{(t-t_0)^2}{4\sigma^2} \right) .  \label{gausspulse} \eeq
Since a Gaussian is infinitely extended we have to start our computer
runs with non-vanishing $\Edach(0)$. We chose $\Edach(0)$ to be 5\% of
the maximum field amplitude $\Edach$. Demanding the Gaussian envelope
to cover $N$ laser cycles within the region $\Edach(t)>0.05\Edach$
yields
\beq t_0=N\pi/\omega, \qquad \sigma^2= t_0^2/(4 \ln 20).\label{gaussparam} \eeq

In Figure \ref{fig_five} the ground state populations for the four
Gaussian pulses with $\Edach=0.3,\ 0.5$ and $N=6,\ 12$ each and
$\omega=0.2$ are shown. Besides, the result for a lower frequency
($\omega=0.1$) and $\Edach=0.5$, $N=12$ is included.
The 12-cycle $\Edach=0.3$-pulse (drawn solid) ionizes most slowly, but the
6-cycle $\Edach=0.3$-pulse (dotted) deplets quicker the ground state than it is
the case for the 12 cycle $\Edach=0.5$-case (dashed). This is due to the fact
that the BSI regime is reached earlier for the weaker but shorter
$\Edach=0.3$-pulse. 

The low frequency pulse (thin solid line) causes more rapid ionization than its
counterpart with twice the frequency since the total time where the
BSI region is reached (measured in {\em absolute} time units) is
larger.

We will further discuss the ground state populations depicted in Figure
\ref{fig_five} in Section \ref{disc} when we reproduce them
with an empirical formula.

\subsection{Sin$^2$ pulses}
In Ref.\ \cite{bauerthesis} one of the authors (D.B.) dealt extensively with
$\sin^2$-pulses of the form
\beq E(t)=\Edach \sin^2\left( \frac{\pi}{T} t\right) \sin\omega t,
\qquad T=N\times \frac{2\pi}{\omega} . \eeq
Since the results look very similar to those in the Gaussian case we
suppress a further discussion here. However, in Section \ref{disc} we
utilize rates numerically determined in \cite{bauerthesis} for
$\sin^2$-pulses in order to confirm the insensitivity of our proposed
rate formula with respect to the pulse shape. Furthermore, a different
numerical scheme was used in \cite{bauerthesis}. This gives additional
reliability to the numerical results which will be utilized to derive
an empirical BSI rate in the following Section.

\section{Discussion} \label{disc}
In this Section we want to demonstrate that it is possible to
reproduce our numerical results using a simple formula for the
ionization rate in the BSI regime. This rate is not sensitive to laser
frequency and pulse shape in a wide parameter range. 
Moreover we show that none of the
analytical rates stated in Subsection \ref{ionirates} is applicable to BSI.

Since BSI occurs mainly during one or two half laser periods a cycle-averaged
rate obviously makes no sense. Therefore the laser field $E(t)$ with
its entire time-dependence has to be plugged in a rate formula, i.e.,
$W(t)=W[E(t)]$, while in tunneling ionization a rate which depends on the
pulse envelope only, $W(t)=W[\Edach(t)]$, is sufficient.

We determined {\em instantaneous} ionization rates from the decreasing
ground state populations, in accordance with Eq.\ \reff{ionirate}. 
In Figure \ref{ratescomp} the results
are plotted vs the electric field present at the corresponding
instant. 
Usually the deepest descent in the ground state population is in the 
vicinity of the electric field maximum of the actual half cycle. 
However, this behavior might be disturbed by ``backsweeping'' 
probability density
ionized earlier, especially for high frequencies (frequencies not much
less than $\vert\energy_0\vert$) since the excursion length of a
freely oscillating electron is then not much larger than the width of
its wave-packet representation. 

In Figure \ref{ratescomp} different symbols are used for
different pulse shapes, pulse lengths, and laser frequencies. For
comparison the predictions by the analytic formulas of Subsection
\ref{ionirates} are drawn as well. The scattering of the numerical
data is due to the fact that instantaneous rates for a certain
electric field value may stem from runs with different pulse shapes,
peak field strengths or laser frequencies. 

The BSI regime for atomic hydrogen sets in for $E=0.146$ when a
classical electron, initially on an $\energy_0=-0.5$ orbit, can escape
from the atomic core. In general this so-called {\em critical field} in
the case of hydrogen-like ions is given by \cite{commentii,shake,bauer}
\beq \Ecrit = (\sqrt{2} +1 ) \vert\energy_0\vert^{3/2} . \eeq
Once the critical field is reached one expects rapid ionization
within a few half cycles. Therefore we are especially interested in the
region where $E\geq 0.15$. Fortunately, the scattering of our
numerical data is
small in this region of field strengths. This makes possible our goal to provide 
a BSI rate formula valid for a wide range of pulse shapes and laser
frequencies.

We observe that none of the analytical theories under consideration
predicts the BSI rates correctly. Apart from Keldysh's result all
formulas overestimate the ionization rate in the region of interest,
$0.15\leq E \leq 0.5$. The ionization rate for much higher field
strengths might be of academic interest since such high field strengths
cannot be reached without strongly 
ionizing the hydrogen atom during earlier parts of the pulse where the
field strength is in the region we focus on in this paper. In real
experiments, with rare gases for instance, there are of course stronger
bound electrons which get free not before $E\gg 0.1$ but for those
electrons $\Ecrit$ is larger too. We will discuss the scaling
behavior of the ionization rate with respect to $Z$ lateron. 

The rates of Posthumus (P) and Mulser (M) saturate at higher field
strenghts. This is owing to taking the {\em unperturbed} inneratomic
motion to derive an ionization current. In reality, however, the
external field influences the inneratomic motion of the electron and
yields a higher ionization current. The tunneling theories (L, A1 and
A2) are even worse when extrapolated to higher field strengths; 
they predict a decreasing ionization rate which is clearly
unphysical. Note that ``stabilization'' cannot occur when ionization
lasts less than one laser cycle. Although the Keldysh rate (K) does not
suffer from these shortcomings it underestimates the ionization rate
by a factor three and more.

The numerically determined ionization rates in the region $0.15\leq E
\leq 0.5$ can be nicely fitted by $W=2.4 E^2$.  Since every realistic
pulse passes through a region where the electric field is within the
tunneling regime we propose a combined formula
\beq W(t) = \left\{ \begin{array}{l}
\mbox{\rm $W'[E(t)]$ \quad for \quad  $E(t) < E'$} \\
\mbox{\rm $2.4 \times E(t)^2$ \quad for \quad $E(t) \geq E'$}  
\end{array} \right. \label{ourformula}  \eeq
where $E'$ is a threshold electric field determined by imposing $W(t)$
to be continuous, and $W'(t)$ is an appropriate tunneling rate. For
the Landau rate $E'=0.084$ holds.

In Figure \ref{fig_seven} the solid curves were calculated by applying
the BSI rate \reff{ourformula} to the four Gaussian pulses which led
to the results already depicted in Figure \ref{fig_five}. For $W'$ we used the Landau tunneling
rate. The agreement with the exact numerical results (drawn dotted) is
satisfactory. Deviations, especially in the $N=12$,
$\Edach=0.3$-run, are mainly due to the (even for lower field
strength) not very accurate Landau rate. For shorter pulses and higher
peak field strengths the agreement becomes excellent. The dashed curve is
the result when the Landau rate alone is applied to the entire $N=6$,
$\Edach=0.5$ pulse; the ionization rate is strongly overestimated.   

In Figure \ref{fig_eight} the BSI rate \reff{ourformula} was
evaluated for the square pulses discussed in Subsection \ref{sqp}. In
the upper plot the agreement with the numerical results for the
$\Edach\sin\omega t$-case is good. However, in the lower plot
($\Edach\cos\omega t$-case) the agreement is not particularly good since the
abrupt jump in the field strength from $0$ (for $t\leq0$) to $\Edach$ for ($t>0$) leads to transient dynamics which cannot be reproduced by our
simple rate \reff{ourformula}. Therefore, care has to be excercised
for laser pulses where the BSI regime is reached rather abruptly on time scales
shorter than one quarter laser cycle. In all other cases the rate
formula \reff{ourformula} worked well.

\subsection{Scaling}
The TDSE \reff{schr_laengen} can be rescaled to the atomic
hydrogen-case by substituting
\beq \tilde{\vektr} = Z\vektr, \quad \tilde{t}=Z^2 t, \quad
\tilde{\omega}=\omega/Z^2, \quad \tilde{\vektE}=\vektE/Z^3. \eeq
Since our BSI rate is not sensitive to $\omega$, and an ionization rate
has the dimension of an inverse time the rescaled result reads
\beq W(t) = \left\{ \begin{array}{l}
\mbox{\rm $W'[E(t)]$ \quad for \quad  $E(t) < E'$} \\
\mbox{\rm $ 2.4/Z^4 \times E(t)^2$ \quad for \quad $E(t) \geq E'$}  
\end{array} \right. \label{ourformularesc}  \eeq

\subsection{The role of the Keldysh parameter}
The Keldysh parameter
\beq \gamma=\left(\frac{\vert\energy_0\vert}{2U_p}\right)^{1/2} \eeq
with $U_p$ the ``ponderomotive potential'' $U_p=E^2/(4\omega^2)$,
i.e., the mean quiver energy of an electron in the laser field, has to be much less
than unity when tunneling theories such as ADK are derived. The
Keldysh parameter has the vivid physical interpretation of tunneling
time measured in units of the laser period. Does the Keldysh parameter
reveals some significance in the BSI regime too? First of all we note that the
Keldysh parameter in our numerical examples is not much less than
unity. In the $\Edach=0.3$, $\omega=0.2$-case it is 0.67, in the $\Edach=0.5$, $\omega=0.1$-case it is 0.2. 
Thus, in 
commonly used terms in this field, we are rather in the
multiphoton than in the (to BSI extended) tunneling regime. 

However, the static field-rates in Fig.~\ref{fig_one} are also well covered by our empirical rate. Additional test runs at intermediate frequencies yielded
 good agreement also.
Thus, the insensitivity of our BSI rate with respect to the laser
frequency (from static fields up to $\omega=0.2$) 
shows that there seems to be no need to put much emphasis on the
concept of the Keldysh parameter in BSI. However, we did not deal with
frequencies $\geq \vert\energy_0\vert$ in this paper. Moreover, a
small laser frequency keeps the portion of already ionized probability
density far away from the ionic nucleus most of the time since the excursion length
is large. Therefore, the ionization curves for smaller frequencies
usually look ``cleaner'' since interference with parts of the wave
function representing the already ionized electron is suppressed.

\section{Conclusion} \label{conclusion}
We conclude that even for the simplest atom we can think of, i.e.,
atomic hydrogen, none of the theories discussed in this paper predict
correctly the ionization rate in short intense laser pulses reaching
the BSI regime. Thus, extrapolation of tunneling theories to BSI is
not permitted. From the numerical results we deduce that a successful
theory should take the influence of the strong laser field on the {\em inneratomic} dynamics into account. For quantum treatments of strong field ionization this means that one must not make the assumption that the initial state (to be plugged into the transition matrix element) evolves in time as if it was unperturbed (as it is usually done in KFR-type theories). In classical theories (such like the one by Posthumus {\em et al.}) this corresponds to taking the effect of the laser field on the bound Kepler-orbits into acount. However, since in either case, quantum or classical, this appears extremely hard to achieve, empirical rates from numerical simulations of strong field ionization are highly desirable and important as an incredient for other simulation codes, e.g.\ in the field of laser-solid interaction \cite{bauersalo,mucoba,cornolti}.

In this paper, an
empirical formula for the BSI rate has been proposed. Our formula is
not sensitive to pulse shapes and laser frequencies in a wide
parameter range, especially when combined with a reliable tunneling
formula for the weaker parts of the laser pulse.

\section*{Acknowledgment}
This work was supported in part by the European Commission through the TMR
Network SILASI (Super Intense Laser Pulse-Solid Interaction), No.\
ERBFMRX-CT96-0043 and by the Deutsche Forschungsgemeinschaft (DFG)
under contract no.\ MU 682/3-1.

\begin{appendix}
\section{Numerical method} \label{app}
Starting point is the TDSE \reff{sg_in_psi}. We follow the line of
Kono {\em et al.}\ \cite{kono} and perform the substitution \reff{substi}
\beq \Phi(\xi,z,t)=\sqrt{\lambda} \xi^{\lambda-1/2}
\psi(\xi^\lambda,z,t), \qquad \xi^\lambda=\rho,\quad z=z, \quad t=t.
 \label{substiapp} \eeq 
The normalization condition for $\Phi(\xi,z,t)$ simply is
\beq \int_0^\infty d\xi \int_{-\infty}^\infty dz\, \vert
\Phi(\xi,z,t) \vert^2 = 1, \label{normal}\eeq
i.e., we have a ``cartesian''-like volume-element $d\xi\, dz$ for the
normalization of $\Phi$. 

With
\beq
H(t)=K_\xi + K_z + V(t), \label{hamil}\eeq
\beq
K_\xi = -\frac{1}{2\lambda^2 \xi^{2\lambda}} \left\{ \xi^2
\pabl{^2}{\xi^2} -2(\lambda-1)\xi\pabl{}{\xi} +\left(
\lambda-\halb\right)^2\right\},\eeq
\beq
K_z = -\halb \pabl{^2}{z^2}, \eeq
\beq
V(t) = -\frac{Z}{\sqrt{ \xi^{2\lambda} + z^2}}
+\frac{m^2}{2\xi^{2\lambda}} + z\Edach(t)\sin(\omega t+\varphi) \eeq
the TDSE for $\Phi(\xi,z,t)$ 
assumes the form
\beq \imagi \pabl{}{t} \Phi(\xi,z,t) =
H(t)\Phi(\xi,z,t). \label{tdse}\eeq
The goal is to solve this TDSE. 

If $\lambda>1/2$ the transformation \reff{substiapp}
implies that $\Phi(0,z,t)=0$ for all times.
We discretize the $(\xi,z)$-space by
\beq \xi_j=j\Delta \xi, \quad j=1,2,\ldots, J,\qquad z_k=(k-K/2)\Delta
z, \quad k=1,2,\ldots,K \eeq
with constant $\Delta\xi$ and $\Delta z$.
While $\lambda=1$ yields the usual cylindrical coordinate system
$\lambda=3/2$ turned out to offer the numerically more appropriate
choice \cite{kono}. This is owing to the proper treatment of the wave function
near the origin when the finite difference-formulas for the first and
second derivatives in the Hamiltonian \reff{hamil} are applied to the
wave function $\Phi$. Note that uniform spacing in $\xi$ corresponds
to non-uniform spacing in $\rho$. For $\lambda > 1$ the $\rho$
grid-width near the origin is smallest while it gets coarser far away
from the origin.

We use 3-point difference-formulas for all derivatives in $K_z$ and
$K_\xi$ and impose as additional boundary conditions 
\beq \Phi(\xi_J,z,t)=\Phi(\xi,z_1,t)=\Phi(\xi,z_K,t)=0. \eeq
In longer runs we apply a filter each time step  which removes
probability density moving towards the boundaries. This is a somewhat
``shabby'' method (similar to ``imaginary potentials'') but proper
``absorbing boundary conditions'' as discussed in \cite{boucke} are not easily
implemented in more than one dimensions. In any case, we always checked our numerical
results upon sensitivity with respect to grid size and spacing.

The time propagation is performed by applying the evolution operator
\beq U(t+\Delta t)=\frac{1}{[1+\imagi \Delta t
A(t_{n+1/2})/2]}\left(\frac{1-\imagi \Delta t B(t_{n+1/2})/2}{1+\imagi \Delta
t B(t_{n+1/2})/2}\right) [1-\imagi \Delta t A(t_{n+1/2})/2] \label{evol} \eeq
with
\[ A(t)=K_z+\halb V(t), \qquad B(t)=K_\xi + \halb V(t) \]
to the discretized representation of $\Phi(\xi,z,t)$.
This is the so-called Peaceman-Rachford method (PR) \cite{koonin}, the alternating
direction-version of the Crank-Nicholson-method for the TDSE in more than one dimension. The evolution operator
\reff{evol} is second order accurate in time and space (as long as the usual
3-point-difference formulas for the derivatives are used). Provided a
non-iterative method for solving the implicit matrix equations
\begin{eqnarray}
(1+\imagi \Delta t B(t_{n+1/2})/2)\Phi^{n+1/2} &=& ( 1-\imagi \Delta t
A(t_{n+1/2})/2) \Phi^n, \\
(1+\imagi \Delta t A(t_{n+1/2})/2)\Phi^{n+1} &=& ( 1-\imagi \Delta t
B(t_{n+1/2})/2) \Phi^{n+1/2}
\end{eqnarray}
is chosen, the method is unconditionally stable.

The stable ground state on our numerical grid was determined by propagating a
``seed function'' in imaginary time, i.e., we substituted $\Delta t
\to -\imagi\Delta t$ in \reff{evol}. Here, renormalization of the wave
function according \reff{normal} after
several time steps is necessary since imaginary time propagation is not unitary. 
Our experience was that during
imaginary time propagation $\Delta t$ had to be sufficiently small for
$\Phi$ converging to the ground state. A typical choice of our
numerical parameters was (both for real and imaginary time
propagation) 
\[ \Delta \xi = \Delta z = 0.1, \quad \Delta t= 0.05, 
\quad J=60, \quad K=1000 . \]

\end{appendix}

\begin{figure}
\caption{\label{fig_one} Ground state population $\Gamma(t)$ vs time for an
instantaneously switched on dc electric field. After a short transient
behaviour (till $\approx 2$ atomic time units) the rates remain
constant in time. The field strengths $\Edach$ as
well as the constant rates $W$ are indicated in the plot.}
\end{figure}

\begin{figure}
\caption{\label{fig_two} Contour plot of 
the probability density $\vert \psi(\xi,z)\vert^2$ 
after $15.5$ atomic time units  for the $\Edach=0.3$-case. The inlet
shows the same situation as a surface plot. The electron escapes in
negative $z$-direction by ``over the barrier''-ionization. However, a
peak remains at the Coulomb singularity.}
\end{figure}

\begin{figure}
\caption{\label{fig_three} The ground state populations in a strong ac
field for two different peak field strengths ($\Edach=0.3$ and $0.5$)  and three
different phases $\varphi$ each. The dotted lines correspond to
$\varphi=0$, i.e., $\Edach\sin\omega t$, the dashed lines are the
$\varphi=\pi/2$-case ($\Edach\cos\omega t$), and the intermediate case
$\varphi=\pi/4$ is drawn dashed-dotted. }
\end{figure}

\begin{figure}
\caption{\label{fig_four} Contour plot of the probability density $\vert \psi(\xi,z)\vert^2$ 
after 1 laser cycle for the $E(t)=0.3\cos\omega t$-case. Owing to
rescattered probability density wave packets have already formed. The inlet shows the
corresponding surface plot of the probability density. }
\end{figure}

\begin{figure}
\caption{\label{fig_five} Ground state populations for hydrogen in a
Gaussian laser pulse covering $N$ cycles within the region where the
electric field is 5\% of the pulse amplitude $\Edach$ (see
formulas \reff{gausspulse} and \reff{gaussparam} for details). 
  }
\end{figure}

\begin{figure}
\caption{\label{ratescomp} Instantaneous ionozation rates vs the
  electric field present at the certain instant during the
  course of the laser pulse. The results have been obtained from
  different pulse shapes and frequencies: (+) $\sin^2$-pulse with
  $\omega=0.2$, (*) $\sin^2$-pulse with $\omega=0.1$, ($\Diamond$)
  instantaneously switched on dc field, ($\bigtriangleup$) Gaussian
  pulse with $\omega=0.2$. The curves are predictions from various
  analytical theories: (L) Landau, (A1) ADK, (A2) to BSI extended ADK,
  (K) Keldysh, (P) Posthumus, and (M) Mulser. The agreement in the
  region $0.15\leq E \leq 0.5$ is poor. The straight line is $W=2.4 E^2$
  which fits the numerical data in this region quite well.
  }
\end{figure}

\begin{figure}
\caption{\label{fig_seven} Comparison of the numerically determined
ground state populations vs time (drawn dotted) with the analytical
predictions by means of the empirical formula \reff{ourformula} (drawn
solid). The dashed curve shows the result for the $\Edach=0.5$,
$N=6$-result when only  the Landau rate \reff{landaurate} is applied
during the entire pulse.   }
\end{figure}

\begin{figure}
\caption{\label{fig_eight} Comparison of the numerical square
pulse-results with the predictions by formula \reff{ourformula}. In
the upper plot (a) the agreement is very good while in the lower plot
(b) formula \reff{ourformula} suffers from the transient
ionization dynamics caused by the abrupt jump in the electric field at
$t=0$.}
\end{figure}


\begin{references}
\bibitem{au} One atomic mass unit $ = m_e= 9.1094\times 10^{-31}$kg; one
atomic charge unit $= e = 1.6022\times 10^{-19}$C; one atomic action
unit $ = \hbar = 1.0546\times 10^{-34}$Js; one atomic length unit $=
0.5292\times 10^{-10}$m (Bohr radius); one atomic energy unit $ =
27.21$eV; one atomic field strength unit $= 5.1422\times 10^{11}$V/m;
one atomic time unit $= 0.024$fs; one atomic frequency (or rate) unit
$= 4.1341\times 10^{16}$s$^{-1}$; one atomic intensity unit $=
3.5095\times 10^{16}$W/cm$^2$. Useful for practical purposes is the
following formula which converts a given field amplitude (given in atomic
units) into intensity (in W/cm$^2$): $I[\mbox{\rm W/cm}^2]=3.51\times
10^{16}\times \Edach^2[\mbox{\rm a.u.}]$.
\bibitem{commentii} Note, that equating the ground state energy level to the maximum of
the barrier formed by the Coulomb potential and the external field
leads to an underestimated critical field strength because the
Stark-shift does not vanish for a classical orbit in general.
\end{references}
\end{document}